\documentclass[aps,pre,twocolumn,superscriptaddress,amsmath,amssymb]{revtex4}

\usepackage{graphicx}
\usepackage{url,txfonts}
\usepackage{multirow,dcolumn}

\begin{document}

\title{Three faces of node importance in network epidemiology: Exact results for small graphs}

\author{Petter Holme}
\email{holme@cns.pi.titech.ac.jp}
\affiliation{Institute of Innovative Research, Tokyo Institute of Technology, Tokyo, Japan}

\begin{abstract}
We investigate three aspects of the importance of nodes with respect to Susceptible-Infectious-Removed (SIR) disease dynamics: influence maximization (the expected outbreak size given a set of seed nodes), the effect of vaccination (how much deleting nodes would reduce the expected outbreak size) and sentinel surveillance (how early an outbreak could be detected with sensors at a set of nodes). We calculate the exact expressions of these quantities, as functions of the SIR parameters, for all connected graphs of three to seven nodes. We obtain the smallest graphs where the optimal node sets are not overlapping. We find that: node separation is more important than centrality for more than one active node, that vaccination and influence maximization are the most different aspects of importance, and that the three aspects are more similar when the infection rate is low.
\end{abstract}

\maketitle

\section{Introduction}

One of the central questions in theoretical epidemiology~\cite{kiss,ps_rmp,hethcote} is to identify individuals that are important for an infection to spread~\cite{lv,perc_salathe_rev}. What ``important'' means depends on particular scenarios---what kind of disease that spreads and what can be done about it. In the literature, three major aspects of importance have been discussed. First, the \textit{influence maximization} problem is to identify the nodes that, if being sources of the outbreak, would maximize the expected outbreak size $\Omega$ (the number of nodes infected at least once)~\cite{Sun2011,kempe2003maximizing}. Second, the \textit{vaccination} problem is to find the nodes that, if vaccinated (or, in practice, deleted from the network), would reduce the expected outbreak size the most~\cite{perc_salathe_rev}. Third, the \textit{sentinel surveillance} problem is to find the nodes that are likely to get infected early~\cite{christakis_fowler_sentinels,Bajardirsif20120289}. These three notions of importance do not necessarily give the same answer to what node is most important. In this work we investigate how the ranking of important nodes for these three aspects differ and why.

In this paper, we evaluate the three aspects of importance with respect to the Susceptible-Infectious-Removed (SIR) disease spreading model~\cite{hethcote,ps_rmp,kiss,mejn:book} on small connected graphs (all connected graphs from three up to seven nodes). The main reason we restrict ourselves to small graphs is that it allows us to use symbolic algebra, and thus exact calculations~\cite{masuda}. In this way we can discover e.g.\ the smallest graph where the three aspects of importance disagree what node is most important; cf.\ Ref.~\cite{brandes_hildenbrand_2014}.  Graphs of seven nodes are still, we argue, large enough to see effects of distance.

Notwithstanding, large networks are important to study and a possible future extension of this work. Ref.~\cite{radicchi} studies the difference between the influence maximization and vaccination problems on (some rather large) empirical networks. They compare the top results of heuristic algorithms to identify influential single nodes, whereas in this paper we will consider the influence of all nodes, and also all sets of two and three nodes. (Ref.~\cite{radicchi} uses a bit different terminology than us---they call important nodes for vaccination ``blockers'' and important nodes for influence maximization ``spreaders''.)

\begin{figure}
\includegraphics[width=0.9\columnwidth]{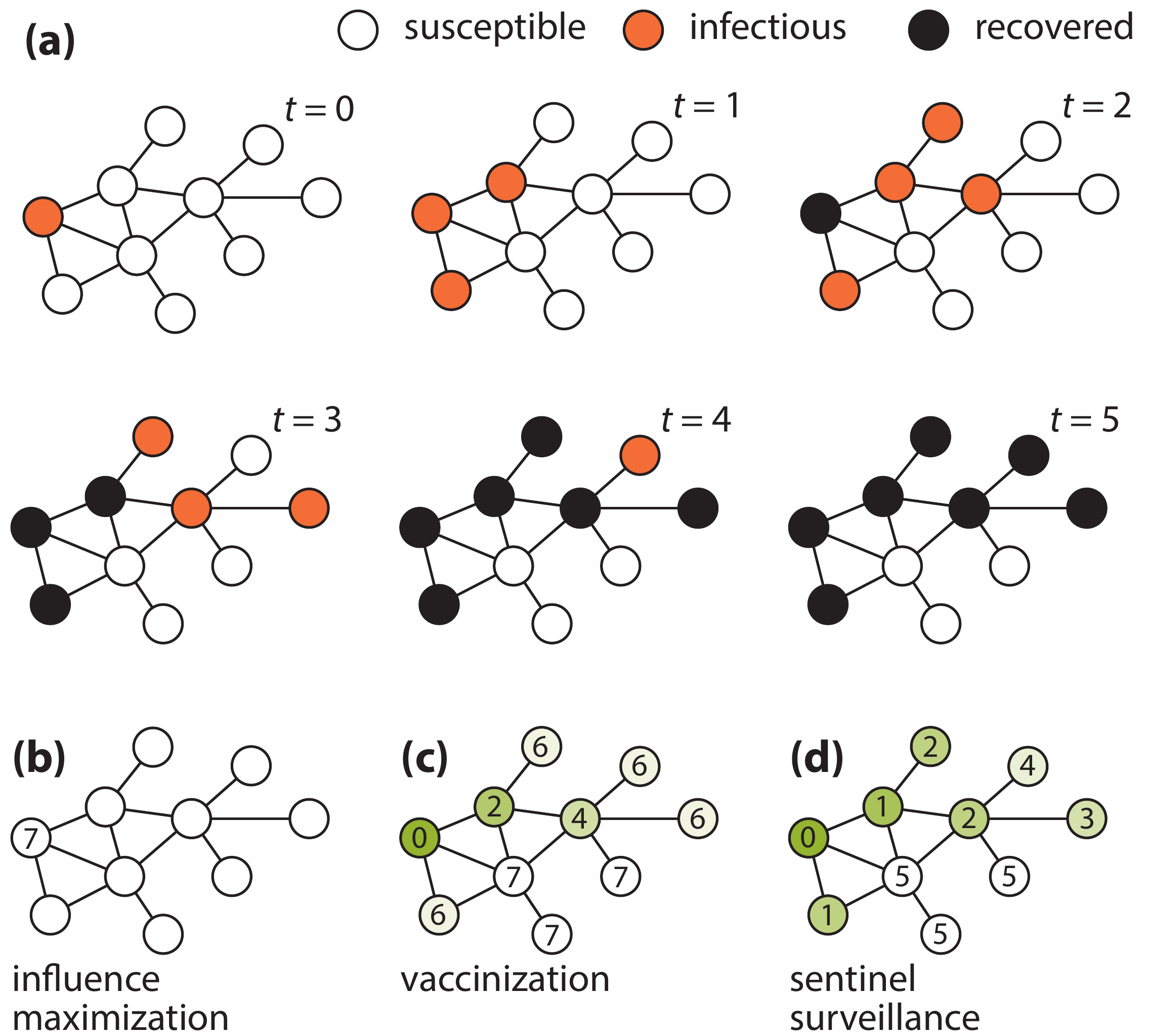}
\caption{(Color online) Illustration the three different notions of importance we explore in this work. Panel (a) shows an example of an SIR outbreak in a 7 node network. Panels (b)--(c) show the contribution of this outbreak to influence maximization (a), vaccination (b) and sentinel surveillance respectively. The idea of influence maximization (b) is that a node is important if it outbreak originating at it are expected to be large. The idea of vaccination (c) is that a node is important if removing it would reduce the average outbreak size much. The idea of sentinel surveillance  (d) is that a node is important if a sensor on it would detect the outbreak early. The shades of the nodes in (c) and (d) is proportional to its contribution. In a stochastic simulation, one would average the values over many runs and, for (c) and (d), many seeds of the outbreak. In this work, however, rather than running simulations, we calculate the exact expectation values of these quantities.}
\label{fig:importances}
\end{figure}

We will proceed discussing our set up in greater detail---our implementation of the SIR model, how to analyze the three aspects of importance, network centrality measures that we need for our analysis and our results including the smallest networks where different nodes come out as most important.

\section{Preliminaries}

In this section, we give the background to our analysis. The basis of our analysis are graphs $G(V,E)$ consisting of $N$ nodes $V$ and $M$ links $E$.

\subsection{Importance}

As mentioned, there are three ways of thinking of importance in theoretical infectious disease epidemiology. Influence maximization was first studied in computer science with viral marketing in mind~\cite{Sun2011,kempe2003maximizing}. As mentioned, a node is important for influence maximization if it being a seed of an infection could cause a large outbreak. For epidemiological applications it could thus be interesting in case one can vaccinate against a disease before an outbreak happens. We will simply measure the expected outbreak size $\Omega(S)$ (the expected number of nodes to catch the disease) with $S$ as the set of source nodes, and rank node set according to their $\Omega$.

For vaccination, we will use the average outbreak size from one random seed node to estimate the importance of a node~\cite{Britton2007graphs,jain,kiss,ps_rmp}. One could, optionally, rephrase it as a cost problem~\cite{nelly}. We assume the vaccinees are deleted from the network before the outbreak starts. The node with smallest $\Omega$ is the one most important for the vaccination problem.

The sentinel surveillance problem assumes a response after the outbreak already started (compared to the influence maximization and vaccination problems where the intervention is assumed to take place before the outbreak happens). A node is important for sentinel surveillance if gets infected early so that the health authorities can activate their countermeasures. This is usually measured by the lead time---the expected difference between the time a sentinel node gets infected, or the outbreaks dies out, and the infection time of any node in the graph~\cite{christakis_fowler_sentinels}. We will instead measure the average discovery time time $\tau(i)$ from the beginning of the infection until a node $i$ gets infected or the outbreak dies~\cite{Bajardirsif20120289}. The node with smallest discovery time is then considered most important for sentinel surveillance. If the purpose of the surveillance is just to discover the outbreak---not to free the population of the disease as early as possible---one could measure $\tau(i)$ conditioned on the outbreak reaching a sentinel before it dies out. We will briefly discuss such a conditioned $\tau$ and refer to it as $\tau'$.

For all of the three problems above, one can consider sets of nodes rather than individuals. There can be more than one source (for influence maximization), vaccinee or sentinel. We will, in general, call these sets \textit{active nodes} and denote the number of them as $n$. We will try to find the optimal sets of active nodes (and call them \textit{optimal nodes}). Note that this is not the same as ranking the nodes in order of importance and take the $n$ most important ones---such a ``greedy'' approach can in many cases fail~\cite{kempe2003maximizing,jain}.

Note that for the vaccination and sentinel surveillance problems, we use one source node of the infection. This is the standard approach in infectious disease epidemiology simply because most outbreaks are thought to start with one person~\cite{hethcote,holme:versions}.

\subsection{SIR model}

We will use the constant infection and recovery-rate version of the SIR model~\cite{holme:versions}. In this formulation, if a susceptible node $i$ is connected to an infectious node $j$, then $i$ becomes infected at a rate $\beta$. Infected nodes recover at a rate $\nu$. Without loss of generality, we can set $\nu=1$ (equivalently, this means we are measuring time in units of $1/\nu$). Let $C$ be a configuration (i.e.\ a specification of the state---S, I or R---of every node); $M_{SI}$ the number of links between S and I nodes; and $N_I$ the number of infected nodes. Then, the rate of events (either infections or recoveries) is $\beta M_{SI}+N_I$ which gives the expected duration of $C$ as 
\begin{equation}\label{eq:dt}
\Delta_t = \frac{1}{\beta M_{SI}+N_I} .
\end{equation}
Proceeding in spirit of the Gillespie algorithm, the probability of the next event being an infection event is $\beta M_{SI}\Delta_t$ and the probability of a recovery event is $N_I \Delta_t$~\cite{ps_rmp,huerta_tsimring}.

\subsection{Exact calculations of $\Omega$ and $\tau$}

Exactly calculating the outbreak size and time to discovery or extinction is, in principle, straightforward. Consider the change from configuration $C$ into $C'$ by an infection event (changing node $i$ from susceptible to infectious). This can happen in $m_i$ ways, where $m_i$ is the number of links between $i$ and an infectious node. Thus the probability for the transition from $C$ to $C'$ is $\beta m_i \Delta_t$. The probability that the next event will be a recovery event is simply $\Delta_t$. To compute the probability of a chain of events one simply multiplies these probabilities over all transitions. To compute the expected time for a chain of events, one sums the $\Delta_t$ for all configurations of the chain.

\begin{figure}
\includegraphics[width=0.9\columnwidth]{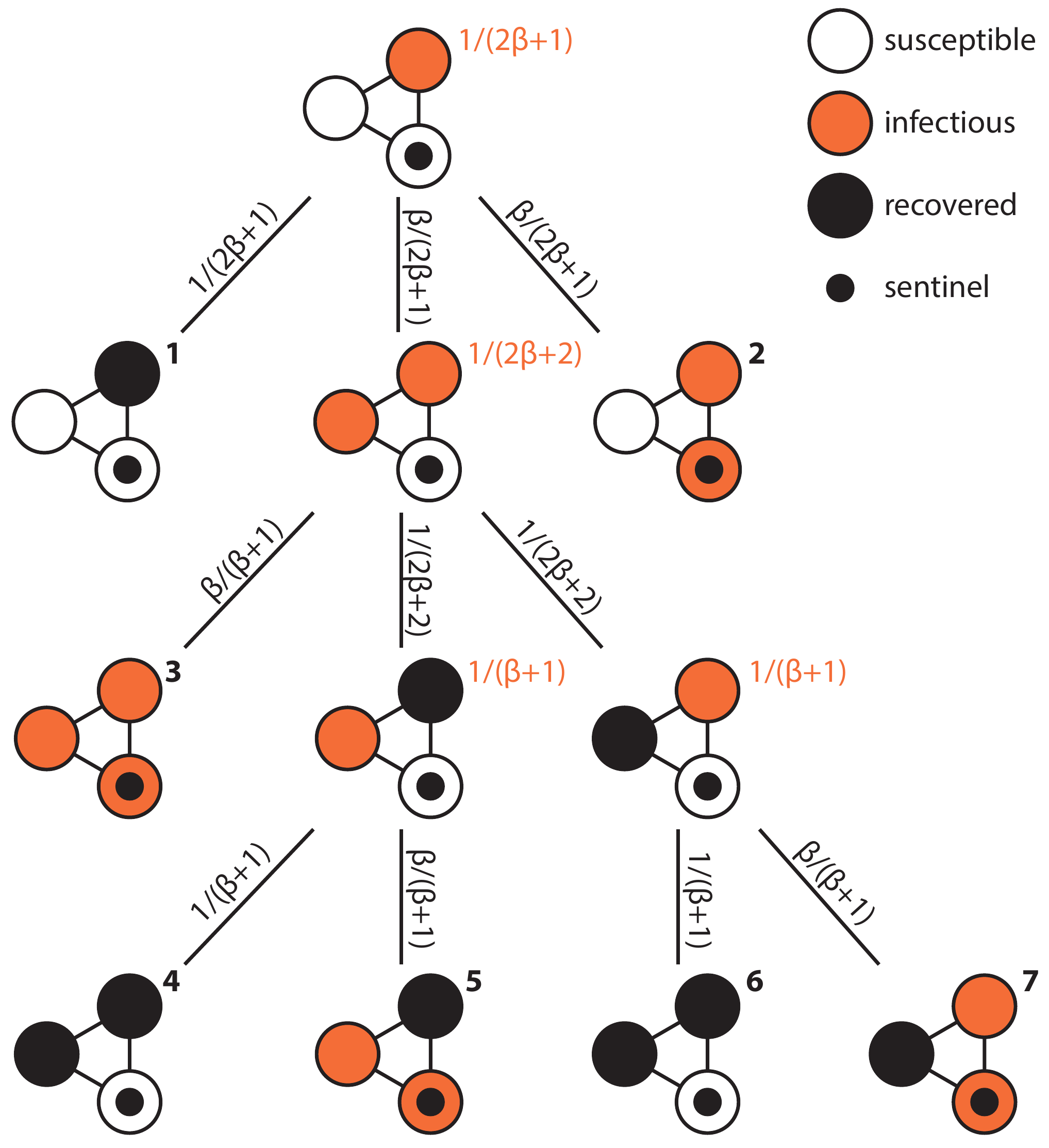}
\caption{(Color online) Illustration of how to calculate the expected time $\tau$ to discovery or extinction in a small network with one sentinel. The probabilities of transitions are marked on the links (with black text). The expected time to stay in a configuration is next to the configurations (in orange). The final states (where there are either no infectious nodes or a sentinel is infected) are enumerated by boldface numbers.}
\label{fig:ill}
\end{figure}

We will illustrate the description above with an example. See Fig.~\ref{fig:ill}. The probability of the outbreak chain 7 is (multiply the probabilities of the transitions)
\begin{equation}
p_7=\frac{\beta}{2\beta+1}\times\frac{1}{2\beta+2}\times\frac{\beta}{\beta+1}=\frac{\beta^2}{2(\beta+1)^2(2\beta+1)}
\end{equation}
The expected duration of the infection chain is
\begin{equation}
\tau_7=\frac{1}{2\beta+1}+ \frac{1}{2\beta+2}+ \frac{1}{\beta+1}=\frac{8\beta+5}{2(\beta+1)(2\beta+1)} ,
\end{equation}
giving a contribution
\begin{equation}
p_7\tau_7=\frac{8\beta^3+5\beta^2}{4(\beta+1)^3(\beta+2)^2} 
\end{equation}
of chain 7 to $\tau$. Then these contributions needs to be summed up for all chains, and averaged over all starting configurations. For the example in Fig.~\ref{fig:ill}, this gives
\begin{equation}
\tau=\frac{3\beta^2 + 7\beta + 2}{6\beta^3 + 15\beta^2 + 12\beta + 3} .
\end{equation}
The expressions of $\Omega$ and $\tau$ are fractions of polynomials. For the largest networks we study (seven nodes), these polynomials can be of order up to 43 with up to 54 digit integer coefficients.

Calculating $\Omega$ for the influence maximization or vaccination problems follows the same path as the $\tau$ calculation above. The difference is that instead of multiplying by the expected time of a chain, one would multiply with the number of recovered nodes in that branch. Furthermore, there are no sentinels to stop outbreaks, so trees (like Fig.~\ref{fig:ill}) become larger.

In practice, our approach to analyzing network epidemiological models is time consuming. The major bottleneck is the polynomial algebra (to be precise---calculating the greatest common divisor, needed to reduce the fractions of polynomials to their canonical form). Because of this, we could not handle more than networks of seven nodes. The code was implemented in both Python (with the SymPy library~\cite{sympy}) and C with the FLINT library~\cite{flint}. It also uses the subgraph isomorphism algorithm VF2~\cite{vf2} as implemented in the igraph C library~\cite{igraph}. Our code is available at \url{http://github.com/pholme/exact-importance}, also including code to calculate $\tau'$ (mentioned above but not investigated in the paper).

\begin{figure*}
\includegraphics[width=\textwidth]{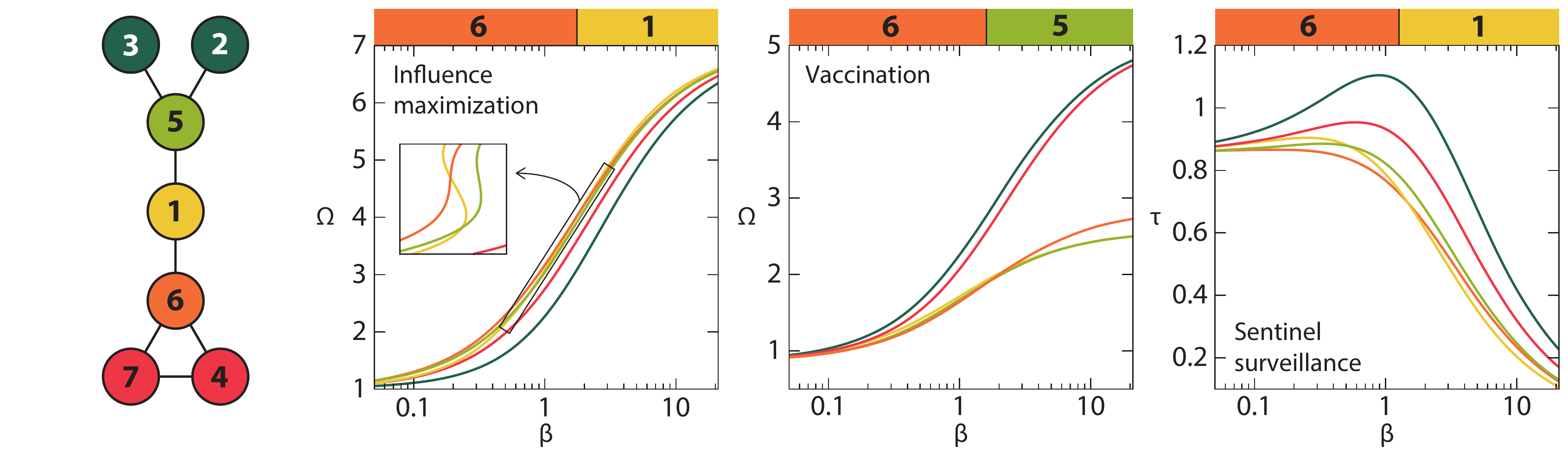}
\caption{(Color online) A small graph with special properties. It is the smallest graph where the three aspects of importance all have distinct nodes as most important. For $\beta$ in the interval $[(1 + \sqrt{5})/2,(3 + \sqrt{17})/4]\approx (1.62,1.78)$, 6 is the most important node for influence maximization, 5 for vaccination and 1 for sentinel surveillance. The curves are color coded according to the graph to the left (with one curve and color per automorphic equivalence class). The inset in panel (a) shows a zoomed and rotated view of some of the crossings of the curves (one of them being the point where node 1 takes over from node 6 as the most important).}
\label{fig:spec}
\end{figure*}

\begin{figure*}
\includegraphics[width=\textwidth]{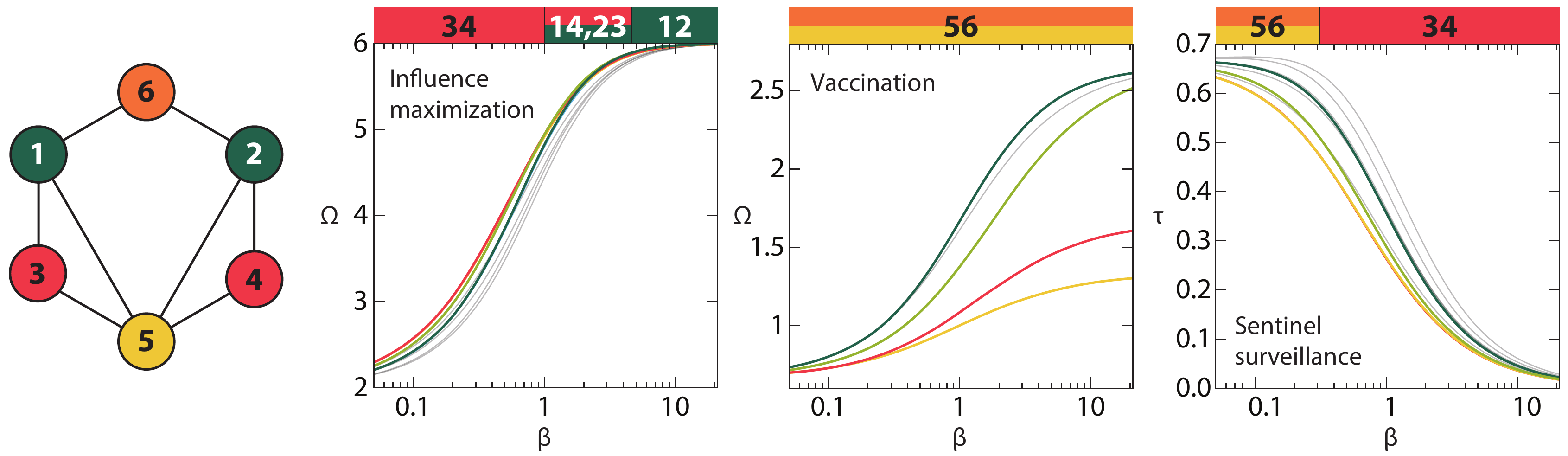}
\caption{(Color online) A special graph for two active nodes. To the left, the colors highlight the automorphic equivalence classes. For $\beta>4.62$, $\{1,2\}$ are the most important source nodes (influence maximization) $\{5,6\}$ are the most important for vaccination and, finally, $\{3,4\}$ are the most important for sentinel surveillance. The color bars indicate the most important node-pairs as a function of $\beta$, the curves are showing the $\beta$ dependence of the objective quantities $\Omega$ and $\tau$. The four node combinations that are optimal at some $\beta$ value for some scenario are highlighted, others are gray.}
\label{fig:spec2}
\end{figure*}

\subsection{Centrality measures}

To better understand how the network structure determines what nodes are most important, we measure the average values of static importance predictors. In general, there can be many ways of thinking of being central means for a node---is it a node that things traveling over the network often passes, or is it a node that have short paths to others, etc. Different rationales give different measures. These are typically positively correlated, but does not rank the nodes in the exactly same way, and can thus complement each other~\cite{koschutzki2005centrality}. We focus on three measures: degree, closeness centrality and vitality.

\textit{Degree} centrality is simply the number of neighbors of a node. If a node has twice the neighbors of another, it has twice as many nodes to spread an infection to. This makes it more important for influence maximization and vaccination. It also has twice as many node to get the infections from, which contributes to its importance for vaccination and sentinel surveillance. On the other hand, degree is not a global quantity---it could happen that the neighbors of a high-degree nodes are so peripheral that a disease could easily die out there. The simplest way of modifying degree to become a global measure is to operationalize the idea that a node is central if it is neighbor of many central nodes. With the simplest possible assumptions, this reasoning leads to \textit{eigenvector centrality}---that the centrality of node $i$ can be estimated as $i$'th entry of the leading eigenvector of the adjacency matrix~\cite{mejn:book}. For the small graphs we consider, however, the eigenvector centrality is so strongly correlated with degree (intuitively so, because ``everything is local'' in a very small graph), so it makes little sense to include it in the analysis.

Many centrality measures are based on shortest paths. Perhaps the simplest being \textit{closeness centrality}---using the idea that a node is central if it is on average close to other nodes~\cite{koschutzki2005centrality,mejn:book}. This leads to a measure of the centrality of $i$ as the reciprocal distance to all other nodes in the network:
\begin{equation}
c(i)=\frac{N-1}{\sum_{j\neq i} d(i,j)}.
\end{equation}
The main problem, in general, with closeness centrality is perhaps that it is ill-defined on disconnected graphs. In our work, however, we consider only connected graphs.

We chose the third centrality measure---\textit{vitality}---with the vaccination problem in mind. Vitality is, in general, a class of measures that estimates node centrality by its impact on the network if it is deleted~\cite{koschutzki2005centrality}. In our work, we let vitality denote the particular metric
\begin{equation}\label{eq:vitality}
v(i) = \frac{s(G)-1}{s(G\setminus \{i\})}
\end{equation}
where $s(G)$ is the number of nodes in the largest component of $G$. This measure is thus in the interval $[1,N-1]$ and increases with $i$'s ability of, if removed, fragmenting the network. Since vaccination is, in practice, like removing nodes from the network, we expect $v$ to identify important nodes for $\beta$ close to one. For large graphs we expect $v$ to be very close to one, so we only recommend it for small graphs such as the ones we use. Another popular centrality measure---\textit{betweenness centrality} (roughly how many shortest paths in the network that passes a node)~\cite{mejn:book}---is very strongly correlated with vitality for our set of small graphs, and thus omitted from the analysis.

\begin{figure*}
\includegraphics[width=0.8\textwidth]{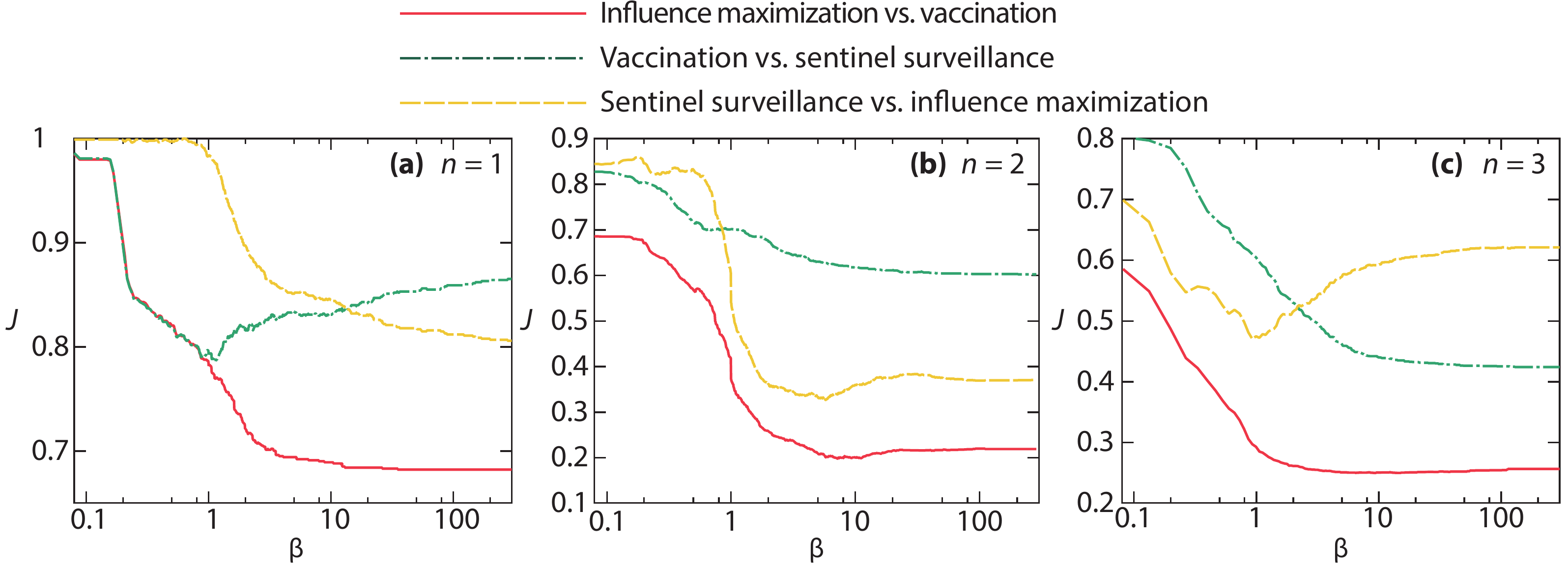}
\caption{(Color online) The average pairwise Jaccard overlap between the sets of most influential nodes for the three different aspects of importance. Panels (a), (b) and (c) show the overlap for one, two and three active nodes, respectively. For $\beta$ larger than shown, the curves are constant.}
\label{fig:jacc}
\end{figure*}

\begin{figure*}
\includegraphics[width=0.8\textwidth]{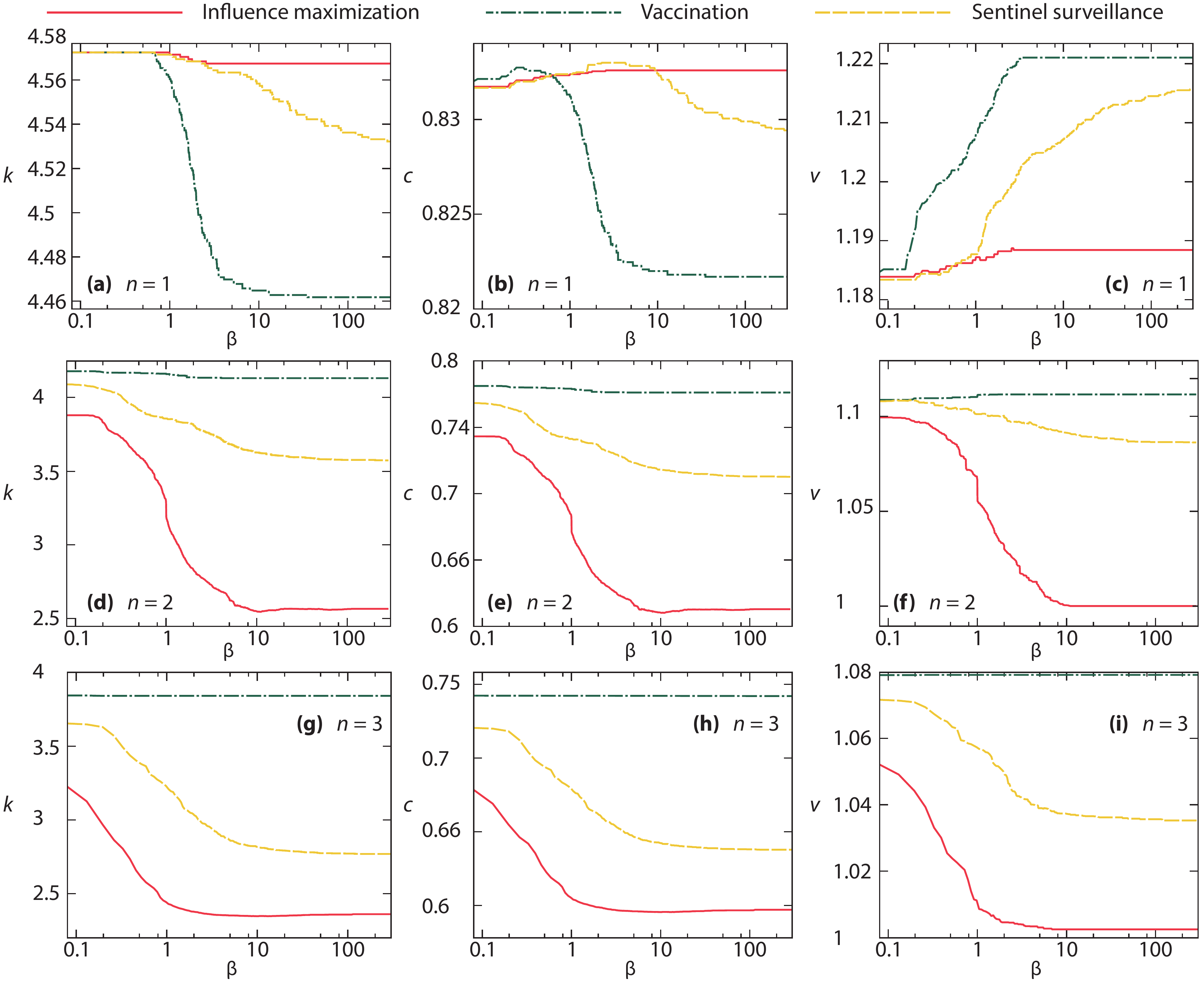}
\caption{(Color online) Average centrality values for the different aspects of centrality and the different numbers of active nodes Panels (a), (d) and (g) show the average degree; (b), (e) and (h) show closeness centrality; and (c), (f) and (i) show curves for vitality. Panels (a), (b) and (c) show results for $n=1$; (d), (e) and (f) for $n=2$; and (g), (h) and (i) for $n=3$.}
\label{fig:centralities}
\end{figure*}

\begin{figure}
\includegraphics[width=0.65\columnwidth]{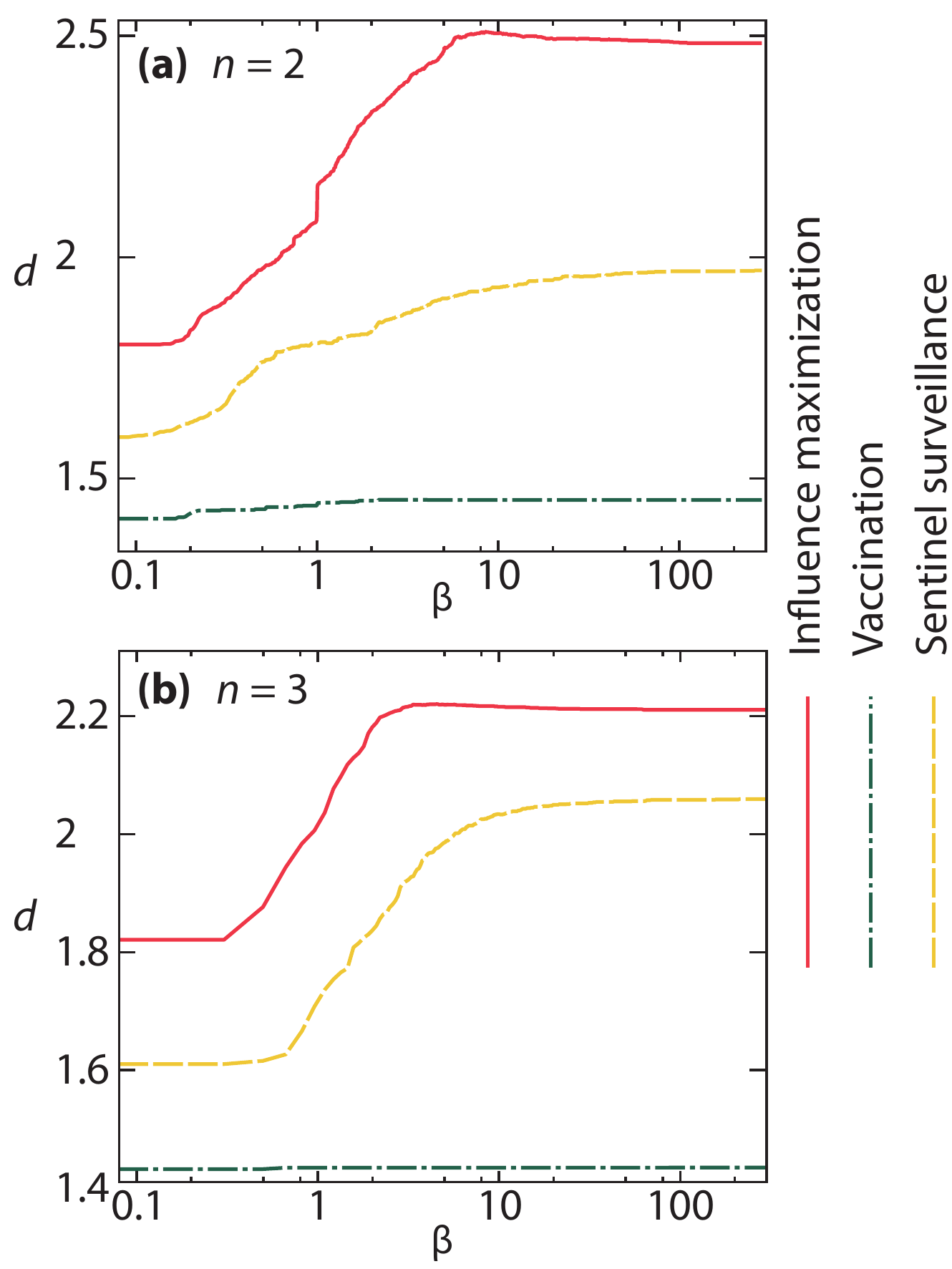}
\caption{(Color online) The average distance between optimal nodes when the number of active nodes are $n=2$ (a) and 3 (b).}
\label{fig:dist}
\end{figure}

\subsection{Small distinct graphs}

In our work, we systematically evaluate small distinct (non-isomorphic) connected graphs. We use all such graphs with $3\leq N\leq 7$. There are two such graphs with $N=3$, six with $N=4$, 20 with $N=5$, 112 with $N=6$ and 853 with $N=7$. To generate these, we use the program Geng~\cite{McKay201494}.

\section{Results}

In our analysis, we will focus one the question when and why the three cases of node importance rank nodes differently. We will start with some extreme examples and continue with general properties of all small graphs. 

\subsection{Special cases}

Inspired by Ref.~\cite{brandes_hildenbrand_2014}, we will start with a special example (Fig.~\ref{fig:spec}). This is the smallest graph where the most important single node ($n=1$) is different for influence maximization, vaccination and sentinel surveillance. For $[(1 + \sqrt{5})/2,(3 + \sqrt{17})/4]\approx (1.62,1.78)$, node 6 is the most important for influence maximization, 5 for vaccination and 1 for sentinel surveillance. For small $\beta$-values, 6 is most important for all three aspects of importance. In this region, the outbreaks die out easily. The fact that 5 and 6 have larger degree than the others is, of course, helpful for an outbreak to take hold in the population. 6 is slightly more important as a seed node since the extra link in its neighborhood helps the outbreak to persist longer (there are the $(6,7,4)$ and $(6,4,7)$ infection paths that, although unlikely, does not exist for diseases starting at 5). This reasoning also explains why 6 is most important for vaccination. For sentinel surveillance and for low enough $\beta$, the outbreak would typically end by the outbreak becoming extinct rather than hitting a sentinel. Thus, for low $\beta$, when an outbreak has the highest chance of surviving if it starts at 6, then putting a sentinel is good because an outbreak is either instantly discovered, or likely soon extinct. With a conditional discovery time $\tau'$, the curves are strictly decreasing (since the early die-off is omitted) so 1 is the most important node for all $\beta$.

For larger $\beta$, node 1 becomes, relatively speaking, more important for influence maximization and sentinel surveillance. This is the most central node in many aspects other then degree (it has the largest $c$ and $v$ values, but is also the most central node in other ways not discussed here). For vaccination, however, node 5 is the most important as it fragments the network most (the vitality is the same for both nodes $v(5)=v(1)=2$, but the size of the second biggest component is larger if 1 is deleted). So since 1 becomes more important than 6 at a larger $\beta$ value for influence maximization compared with sentinel surveillance, there is an interval of beta where the network of Fig.~\ref{fig:spec} has three distinct most important nodes for the three aspect of importance we investigate.

For two active nodes ($n=2$), the smallest network with no overlap between the optimal node sets is actually smaller than for $n=1$. This network, displayed in Fig.~\ref{fig:spec2}, has six nodes and eight links. Note that $N=6$ is the smallest number of nodes to make three distinct sets of two nodes, so in that sense the $n=2$ example seems more extreme than the $n=1$ one of Fig.~\ref{fig:spec}. 

For large $\beta$ values, 1 and 2 are the most important nodes for influence maximization, 5 and 6 are most important for vaccination, 3 and 4 are most important for sentinel surveillance. 5 and 6 are the nodes that, if deleted, break the network into smallest components, which explains why they are most important for vaccination (at least for large $\beta$). In addition to 5 and 6, 1 and 2 is the only pair of nodes whose neighborhoods contain all other nodes. 1 and 2 have both degrees 3, as opposed to 5 and 6 that have degrees 4 and 2 respectively. If and why that makes 1 and 2 better than 5 and 6 for influence maximization is not clear. Similarly it is hard to reason about why 3 and 4 are the best nodes for sentinel surveillance. The neighborhoods of these nodes do not even contain the entire graph.

We can see that the optimal sets of nodes in Fig.~\ref{fig:spec2} do not have links within themselves. This seems natural for most networks and all three notions of importance. This means that as $n$ grows, the distance between the optimal nodes will be larger than one. This is an observation we will make more quantitative in the next section. Another such observation is that for small $\beta$ the optimal nodes for the three importance aspects are overlapping. In this parameter region, most outbreaks dies out before they reach a sentinel. If the outbreak starts at a high-degree node in a highly connected neighborhood, there is a larger chance for it to survive. For all three importance aspects, it is important to have active nodes where an outbreak would be likely to survive. Still, as evident from Fig.~\ref{fig:spec2}, there are examples where the optimal nodes are not overlapping.

\subsection{$\beta$-dependence}

Now, we will move to a more statistic evaluation of all graphs with 3 to 7 nodes. We will present average quantities over all these graphs as functions of $\beta$. Other summary statistics, including grouping the graphs according to size, give the same conclusions.

Let $u^{a,b}_i$ be the optimal sets for a given network, $\beta$ and importance classes $a$ and $b$. The first quantity we look at is the pairwise overlap of sets of optimal nodes as measured by the Jaccard overlap
\begin{equation}
J(a,b)=\frac{|U_a \cap U_b|}{|U_a\cup U_b|}
\end{equation}
where
\begin{equation}
U_{a,b} = \bigcup_i u^{a,b}_i .
\end{equation}
For example, in Fig.~\ref{fig:spec2} at $\beta=2$, we have:
\begin{subequations}
\begin{eqnarray}
u^a_1&=&\{1,4\}\\
u^a_2&=&\{2,3\}\\
u^b_1&=&\{3,4\},
\end{eqnarray}
\end{subequations}
where $a$ is influence maximization and $b$ represents sentinel surveillance, giving
\begin{equation}
J(a,b)=\frac{|\{1,2,3,4\}\cap\{3,4\}|}{|\{1,2,3,4\}\cup\{3,4\}|}=\frac{1}{2}.
\end{equation}

We show the average Jaccard similarity between the different aspects of importance in Fig.~\ref{fig:jacc}. The general trend is that $J$ decreases from an early maximum, i.e.\ the overlap between the different aspects of importance is highest when $\beta$ is smallest. For $n=1$, $J$ is close to one in the limit of small $\beta>0$. (For $\beta=0$, all node are equally important in all three senses of importance.)

For $n=1$, however, the overlap between the optimal nodes for vaccination and sentinel surveillance has a minimum as functions of $\beta$. The same is true for sentinel surveillance versus influence maximization when $n=3$. It is hard to say why, more than that, for individual graphs, the $J(a,b,\beta)$ curves can of course be non-monotonous as different aspects of the graph structure determine the role of the nodes. We note that (for a different spreading model and much larger networks), Ref.~\cite{radicchi} finds the Jaccard similarity between influence maximization and vaccination to have a minimum as a function of $\beta$.

\subsection{Structural predictors}

Next, we investigate the structural properties of the most influential nodes and how they depend on $\beta$. In Fig.~\ref{fig:centralities}, we plot the degree, closeness centrality and vitality as a function of $\beta$ for all aspects of importance and $n\in\{1,2,3\}$. We start examining the case $n=1$, Fig.~\ref{fig:centralities}(a), (b) and (c). A first thing to notice is that the general impression is that centralities of the optimal nodes decrease with $\beta$. The only case with an opposite trend is vitality (Fig.~\ref{fig:centralities}(c)), where the curves are monotonically increasing. If we first focus on the case with one active node, this could be understood as the ability of nodes to (if removed) fragment the network. This ability is captured by vitality and becomes more important as $\beta$ increases.

Continuing the analysis for $n=1$, when $\beta$ is low the most important thing is for the outbreak to persist in the population. If an active node has a high degree, it is likely to be the source of a large outbreak, meaning it is important for influence maximization (which was also concluded by Ref.~\cite{radicchi}). If a high-degree node is deleted it would remove many links that could spread the disease and thus be important for vaccination~\cite{ps_vesp}. It would also be important not to put a sentinel on a low-degree node for sentinel surveillance and low $\beta$ as diseases reaching low-degree nodes would be likely to die out. So panels Fig.~\ref{fig:centralities}(a) and (c) can be understood as a shift from nodes of high degree to nodes of high vitality. Closeness centrality---seen in Fig.~\ref{fig:centralities}(b)---is harder to explain. Values of $c$ increases with $\beta$ for influence maximization but decreases for vaccination. One way of understanding this is from the observation that vitality is most important for vaccination (as evident from Fig.~\ref{fig:centralities}(c)), and degree is most important for influence maximization (as seen in Fig.~\ref{fig:centralities}(a)). The results of Fig.~\ref{fig:centralities}(b) then suggests that the high vitality nodes optimizing the solution of the vaccination problem have a lower closeness centrality. Indeed, for many of the graphs we study, the highest vitality node has many degree-one neighbors---cf.\ node 5 in Fig.~\ref{fig:spec}---which is not necessarily contributing to the closeness centrality. For influence maximization, it seems that the optimal nodes are central in the closeness sense---the closer to average the seed node is to the rest of the network, the higher the chance is for the outbreak to reach all the network.

For $n=2$ and 3, the picture is somewhat different than for $n=1$. In these cases, all centrality measures are monotonically decreasing. The order of the importance measures are all the same with vaccination having the largest values, and influence maximization the smallest. It is no longer the case for vaccination that the optimizing nodes have high vitality and low closeness centrality (as it was for $n=1$. Indeed, for the vaccination case, the optimal nodes are usually independent of $\beta$ which is why the curves for vaccination in Fig.~\ref{fig:spec}(d)--(i) are almost straight. Naively, one would think that some centrality measure needs to increase with $\beta$. However, as we will argue further below, the optimal nodes would usually not be be close to each other. One could think of each node being responsible for (and centrally situated within) a region of the network, and that that tendency is so strong that it overrides all simple centrality measures. On the other hand, there are group centrality measures that could perhaps increase with $\beta$~\cite{group} (that could be a theme for another paper).

\subsection{Distance between optimal nodes}

The fact the all curves of Fig.~\ref{fig:centralities}(d)--(i) are non-increasing could be explained by that the separation of the optimal nodes increases with $\beta$. 
In Fig.~\ref{fig:dist}, we try to make this argument more quantitative by measuring the average (shortest path) distance $d$ between the optimal nodes. In the limit of small $\beta$, these values comes rather close to its minimum of $1$, but as $\beta$ increases, so does $d$. Essentially, the pattern from Fig.~\ref{fig:dist} is the reversed of Fig.~\ref{fig:centralities}(d)--(i)---the vaccination curve is almost constant, sentinel surveillance increases moderately but influence maximization increases much more.

A larger separation gives the sentinels the ability to on average be closer to outbreaks anywhere in the network, while for influence maximization a larger separation means that there are more susceptible-infectious links (less infectious-infectious links) in the incipient outbreak. For vaccination there is no such positive effect of a larger separation that we can think of, which is a part of the explanation why the optimal sets are relatively independent of $\beta$ for $n>1$. The rest of an explanation---why the trends for $n=1$ are so much weaker when $n>1$ is not clear to us, and something we will investigate further in the future.

\section{Discussion}

Furthermore, we investigated average properties of the optimal nodes for all our graphs. We found that the overlap between the optimal nodes of the different importance aspects are largest for small $\beta$. In the small-$\beta$ region, a high degree seems most important for all importance aspects. For larger $\beta$ nodes positioned such that they would fragment the network much if they were removed becomes more important, in particular for the vaccination problem (slightly less for the sentinel surveillance problem, and much less for influence maximization). When the number of active nodes, on the other hand, increases, it becomes important for the nodes to be spread out---the average distance between them increases. This effect is large for influence maximization, intermediate for sentinel surveillance, and very small for vaccination. The small effect for vaccination can be understood since all that matters is to fragment the network, and for that purpose the vaccinees does not necessarily have to be distant.

Most of the behavior discussed above seems quite natural. For small $\beta$, the dominant aspect of the dynamics is how fast an outbreak will die out. For large $\beta$, the outbreak will almost certainly reach all nodes. For vaccination and sentinel surveillance, this leads to a question of deleting nodes that would break the network into smallest components. (In the former case, this is trivial since the size of the outbreak almost surely is the size of the connected component to which the seed node belongs. In the latter, we conclude this from the monotonically increasing vitality.) 

To extend this work, it would be interesting to confirm the picture pained above for larger networks using stochastic simulations. This would not allow the discoveries of special graphs such as those in Figs.~\ref{fig:spec} and \ref{fig:spec2}, but could put the connection between the different notions of centrality on a more solid footing. We believe many of our conclusions hold for larger networks, an indication being that our results are consistent with the results of Ref.~\cite{radicchi} (comparing the vaccination and influence maximization for $n=1$ in large empirical networks).

\begin{acknowledgments}
We thank Petteri Kaski, Nelly Litvak and Naoki Masuda for helpful comments.
\end{acknowledgments}

\bibliographystyle{abbrv}
\bibliography{exact}

\end{document}